\documentclass[lettersize,journal]{IEEEtran}
\usepackage{amsmath,amsfonts}
\usepackage{algorithmic}
\usepackage{algorithm}
\usepackage{array}
\usepackage[caption=false,font=normalsize,labelfont=sf,textfont=sf]{subfig}
\usepackage{mathtools}
\usepackage{textcomp}
\usepackage{stfloats}
\usepackage{url}
\usepackage{verbatim}
\usepackage{graphicx}
\usepackage{cite}
\usepackage{amsmath,amssymb,amsfonts}
\usepackage{algorithmic}
\usepackage{graphicx}
\usepackage{textcomp}
\usepackage[numbers,sort&compress]{natbib}
\usepackage{color}
\usepackage{multicol}
\usepackage{bigints}
\usepackage{xcolor}
\usepackage{amsmath} 
\usepackage{amssymb}  
\newtheorem{theorem}{Theorem}
\newtheorem{remark}{Remark}
\begin{document}

\title{\textcolor{black}{Online Detection and Mitigation of Robust Zero Dynamics Anomaly Behavior in MIMO Nonlinear Control Systems}}

\author{Kosar~Behnia,~H.A.~Talebi,~\IEEEmembership{Senior~Member,~IEEE},~and
	~Farzaneh~Abdollahi,~\IEEEmembership{Senior~Member,~IEEE}
	  \vspace{-1.0cm}
	\thanks{Authors are with the Department of Electrical Engineering, Amirkabir University of Technology, Tehran, Iran,
		e-mails: (kosar.behnia@aut.ac.ir, alit@aut.ac.ir, and f$\_$abdollahi@aut.ac.ir).}
	
	\thanks{\textit{Corresponding Author}: H.A.~Talebi (alit@aut.ac.ir)}
}



\maketitle

\begin{abstract}
\textcolor{black}{This paper presents a methodology to detect robust zero dynamics anomaly behavior and mitigate the impacts in general multi-input multi-output (MIMO) nonlinear systems. The proposed method guarantees the resiliency and stability of the closed-loop system without relying on an accurate dynamical model. The presented method operates in two stages. First, it measures the difference between the system input and that of the model as a residual signal to detect the anomaly behavior. After detecting the attack, a recovery signal is generated to restore the system to its nominal condition. In this stage, a neural network model is used to estimate the anomaly signal and recover the closed-loop system. The weights of the neural network model are updated online using adaptation rules without needing prior data for training. The accuracy and performance of the proposed methods are verified by simulating various scenarios on a four-tank system.}

\end{abstract}


\section{Introduction}
\label{sec:introduction}
\IEEEPARstart{U}{S}e of communication \textcolor{black}{networks} in control systems is increasing due to their numerous benefits, such as reducing economic costs and improving data transfer speeds. Unfortunately, in addition to the countless benefits of communication links, the usage of these links makes control systems vulnerable to cyber attacks. \textcolor{black}{The occurrence} of cyber attacks in a control system can have extreme consequences, such as human and financial losses.
\\\indent In recent years, various cyber attacks \textcolor{black}{have occurred in different industries and sectors, such as the BlackEnergy malware that damaged Ukraine’s power grid in 2015 \cite{attack}. Unfortunately, these attacks have caused various damages, such as downtime, data and money loss.} Hence, much attention has been paid to \textcolor{black}{the study of cyber attacks and methods for improving the resiliency of systems.} As a result, \textcolor{black}{various research works have studied different types of cyber attacks, such as covert attacks \cite{de2017covert}, denial of service attacks \cite{long2005denial}, zero dynamics attacks \cite{shim2022zero}, and robust zero dynamics attacks \cite{park2019stealthy} to increase the security and robustness of control systems.}
\\
\indent \textcolor{black}{Zero dynamics attacks are among disruptive cyber attacks causing instabilities and catastrophic damages.} In a zero dynamics attack, the attacker makes the system's zero dynamics unbounded by simply adding a zero dynamics directed signal to the plant's input. In zero dynamics attack, the system's zero dynamics become unbounded while the system output follows the desired trajectory. Therefore, detecting the occurrence of zero dynamics attacks is challenging and cannot be done by monitoring the system output.
\\
\indent An exact \textcolor{black}{model of the control system} should be available to perform a zero dynamics attack. However, the attacker cannot obtain an accurate  model in many practical cases. To overcome this issue, authors in \cite{Uncertain_nonlinear} extended the zero dynamics attack for uncertain nonlinear systems and proposed the robust zero dynamics attack. In this attack, only a nominal model of the system is required, \textcolor{black}{which can be different from the actual dynamics.} Similar to zero dynamics attacks, robust zero dynamics attacks can also cause the system's zero dynamics to be unbounded while having negligible effects on the system output.\\ 
\indent  Many strategies have been proposed to secure and stabilize cyber physical systems subjected to different types of stealthy attacks, specifically zero dynamics attacks. A typical method to detect zero dynamic attacks is to change system representation. This method, presented in \cite{shames}, only applies to linear systems. Hence, this strategy cannot detect robust zero dynamics attacks. Authors in \cite{dcservo} proposed a data driven attack detection method. This method first uses a residual signal to detect \textcolor{black}{zero dynamics attacks} on actuators. Then $H_\infty$ and  $H_-$ indices are used to investigate the robustness of the residual signal. But, the proposed method is only applicable for linear systems. \textcolor{black}{ Authors in \cite{frequency-constrained} proposed a method to detect false data injection attacks using observer based abnormal detectors and adaptive filters. \textcolor{black}{However, the main drawback of this strategy is that it only works for attack signals with bounded amplitudes.} Authors in \cite{encode} presented a method to detect attacks for linear systems by encoding sensors' data and designing a Gaussian anomaly detector on the controller side. \textcolor{black}{Another strategy to detect zero dynamics attacks for linear systems is through measuring a backward-in-time detection residual \cite{passive}. This strategy uses adaptive thresholds to detect zero dynamics attack. Similar to previous studies, this method is designed for linear systems.} Authors in \cite{combine} developed a method based on combining multiple linear regression and predictive control models to protect linear systems against zero dynamics attacks.}\\ 
 \indent \textcolor{black}{Cyber security of DC microgrids is the main subject of the study in \cite{DC-observer}.} The proposed strategy is based on designing Luenberger like observers and generating a residual signal decoupled from unknown loading conditions and neighbor voltage changes. The approach assumes a linear structure for the DC microgrid system. Moreover, an exact system model is required to implement the proposed method, which cannot be obtained in many cases. Therefore, the presented technique is inefficient for detecting robust zero dynamics attacks in DC microgrids with nonlinear components. Authors in \cite{zdfilter} suggested a strategy based on filtering to detect zero dynamics attacks in linear systems. An approximate linear model of the system is needed to use the proposed method for nonlinear systems. However, linearizing a nonlinear system around an operating point \textcolor{black}{results in a linear model which is valid only for the specific operating point. Systems generally have many operating points, especially when the reference signal is time varying. Therefore, this method loses its performance once we change the operating point.} In this regard, different filters based on the operating points of the nonlinear system should be designed, which leads to a high computational cost.\\
\indent In \cite{sampledata,Zd-Turbine}, authors proposed different strategies for stabilizing linear control systems' zero dynamics to reduce the effects of the zero dynamics attacks. For nonlinear systems, linearization error significantly decreases the performance of the proposed methods. More specifically, authors in \cite{Uncertain_nonlinear} show the stealthy property of the zero dynamics attack cannot be guaranteed for a linearized model of a nonlinear system. Moreover, the uncertainty of the underlying nonlinear model can dramatically increase linearization error. Hence, detection and recovery strategies that assume a linear model for the system cannot be used to detect robust zero dynamics attacks.
\\
\indent In recent years, the effectiveness of artificial intelligence (AI) algorithms in control systems has been proven in many studies. Various classification methods based on \textcolor{black}{neural network models can} identify adversarial cyber attacks.  Algorithms based on artificial intelligence usually extract appropriate features in the learning phase that separate malicious data obtained from the system under attack from the data of normal operations. Then, using the collected data and extracted features, different AI structures such as neural network \cite{deep} and random forest \cite{random_forest} can be used to detect cyber attacks. The proper performance of AI based algorithms depends on the learning data. Learning data for many attacks, especially new ones, are not widely available, making the proposed AI method ineffective.
\\
\indent \textcolor{black}{According to the existing studies in the literature, to the best of the authors' knowledge, this study is the first work presenting a strategy for both detection and mitigation of robust zero dynamics attacks in MIMO nonlinear systems. Previous works have applications limited to linear systems and need some restricting assumptions such as having constant reference signals. The proposed method detects the occurrence of a robust zero dynamics attack in the system by observing a residual signal obtained from the system input. Then, in case an attack is detected, a neural network model is obtained through online adaptation algorithms. The neural network model is used to estimate the attack signal and generate the recovery signal for the closed-loop system. The neural network model does not need any prior data and is updated online without any limiting constraints. The estimated signal is then subtracted from the actual system's input to restore the attacked system to its stable operating condition. In summary, the main contributions of this paper can be expressed as follows:}
	\begin{enumerate}
			\item  \textcolor{black}{A detection and mitigation strategy for robust zero dynamics attacks in general MIMO nonlinear systems is developed. This strategy does not need an exact model to detect and recover the system and is independent of prior data to update the neural network model and its weights. Moreover, unlike previous studies, there is no need to linearize the system or consider restricting assumptions on the model or type of attack.}
			\item \textcolor{black} {The mitigation strategy recovers the system to its normal operating point without causing any performance issues for the system and its control objective. The recovery strategy guarantees asymptotic stability of the closed-loop system subjected to attack.}
			\item \textcolor{black}{There is no impractical assumption or limitation on the structure of the nonlinear system, making the proposed method applicable to a wide range of control systems with nonlinear dynamics. Furthermore, this algorithm works in parallel with the system's controller, without the need to modify the system or impose any restrictions.}
		\end{enumerate}

\indent Section \ref{problemstate} states the problem and describes the main properties of robust zero dynamics attacks. Section \ref{detection} describes the attack detection algorithm. Section \ref{recovery} provides details of the recovery strategy. Section \ref{simulation} presents simulation results of the proposed methods on a sample MIMO
nonlinear system. Finally, Section \ref{conclusion} concludes the paper.
\section{Problem statement \label{problemstate}}
\textcolor{black}{In a zero dynamics attack, the attacker injects a signal into the plant input that aligns with the system's zero dynamics directions.
}. If the attack signal is not in the direction of the zero dynamics of the system, the attack can be detected by monitoring the output signal. \textcolor{black}{In practice, the uncertainties inherent in the system's structure make it impossible to determine the exact direction of its zero dynamics. Consequently, the traditional zero dynamics attack becomes less effective.} In \cite{Uncertain_nonlinear}, zero dynamics attacks are extended to be effective for uncertain systems. The following is a brief review of the proposed study in \cite{Uncertain_nonlinear}.
\\
\indent Let's consider the following standard structure for a nonlinear system:
\begin{align}
\label{nonlinearplant}
\dot{z} &= H_{\sigma}(z,x)
\\ \nonumber
\dot{x} &= F_{\sigma}(z,x) + G_{\sigma}(u_c + \alpha)
\\\nonumber
y &= x \nonumber
\end{align}
Here $z \in \mathbb{R}^{n_z}$ and $x \in \mathbb{R}^{n_x}$ are state variables of the system. \textcolor{black}{ $z$ does not appear in the system's output, therefore it can be considered as zero dynamics of the system.} Matrices $G_\sigma, F_\sigma$, and $H_\sigma$ represent system matrices. Signals $u_c \in \mathbb{R}^{n_u}$ and $y \in \mathbb{R}^{n_y}$ represent the controller output and system output, respectively. The robust zero dynamics attack signal is considered as  $\alpha \in \mathbb{R}^{n_{u}}$, \textcolor{black}{which is applied to the system through the input. The input of the plant can be considered as $u_c+ \alpha$. In the case there is no attack, the input of the plant is the same the controller output $u_c$.}
\\
\indent For the system described in (\ref{nonlinearplant}), the robust zero dynamics attack can be modeled as:
\begin{align}
\label{nonlinearattack}
\dot{\tilde{\delta}} &= H_{n}(\tilde{\delta} + z^*_{n} , x^*)
\\ \nonumber
\alpha &= - G_n^{-1}(F_n(\tilde{\delta} + z^{*}_{n} , x^*) - F_n(z^*, x^*))
\end{align}
 with $x^* \in \mathbb{R}^{n_x}$ and $z^* \in \mathbb{R}^{n_z}$ as the steady state values of the system states in the absence of the attack. The steady state value of the system nominal model's zero dynamics is expressed as  $z^*_{n} \in  \mathbb{R}^{n_z}$. Matrices $G_n$, $F_n$, and $H_n$ are the nominal values of the system matrices. \textcolor{black}{The nominal values can be different from the values of the matrices describing the system model in (\ref{nonlinearplant}), which means model and parametric uncertainties are considered in the system. It is worth mentioning that stochastic noise and external disturbances can also be considered as uncertainty in the system's structure. Therefore, the model can include structural and parametric uncertainties, as well as external disturbances.} 
 \\\indent The structure of the system controller, or the controller used for the actual system, is  considered as follows:
\begin{align}
\label{contnon}
&\dot{s} =M(x, y_{\text{ref}}, s)
\\ \nonumber
&u_c = K(s, x)
\end{align}
Here, $s \in \mathbb{R}^{n_c}$ and $u_c \in \mathbb{R}^{n_u}$ are the controller states and controller output, accordingly. The reference signal is considered as $y_{\text{ref}} \in \mathbb{R}^{n_{r}}$ \textcolor{black}{without having any limitations on its properties.} Controller matrices are expressed by $M$ and $K$.
\\\indent
In \cite{Uncertain_nonlinear}, it is shown that if there are a Lyapunov function such as $\mathcal{V}_{\sigma}$ and constants $c_i>0 $ with $i \in \left\lbrace 1,2,3,4\right\rbrace $ for the closed-loop structure described by (\ref{nonlinearplant}) and (\ref{contnon}),  such that (\ref{lypassump}) holds, then for a 
given $ \epsilon > 0$ in the presence of the robust zero dynamics attack described in (\ref{nonlinearattack}), system's output satisfies $\parallel y - y_{\text{ref}} \parallel < \epsilon$ until the attack succeeds. \textcolor{black}{Let's consider $\epsilon$ as a design parameter.} Therefore, by choosing a small value for $\epsilon$, the robust zero dynamics attack will have negligible effects on the system's output. The attack succeeds when the zero dynamics of the system converge to $\tilde{\delta}$, described in (\ref{nonlinearattack}) with a bounded error caused by the attack signal.
\begin{align}
\label{lypassump}
c_1 \parallel [z, x, s] \parallel ^2 &\leq \mathcal{V}_{\sigma}( z, x, s)
\\ \nonumber
\mathcal{V}_{\sigma}( z, x, s) &\leq  c_2 \parallel [z, x, s] \parallel^2
\\ \nonumber
\frac{d\mathcal{V}_{\sigma}}{d(z, x, s)} ~ \mathcal{F}  &\leq  -c_3\parallel [z, x, s] \parallel ^2
\\ \nonumber
\parallel \frac{d\mathcal{V}_{\sigma}}{d(z, x, s)} \parallel &\leq  c_4 \parallel [z, x, s] \parallel
\end{align}
Here matrix F is defined as follows:

\begin{align}
	\mathcal{F} &= \begin{bmatrix} 
	H_{\sigma} (z, x) \\ F_{\sigma} (z, x) + G_{\sigma}(u^*_{c} + K(s,x)) 
	\\ \nonumber
	M(x, y_{\text{ref}}, s)
	\end{bmatrix} 
\end{align}
where  $u^*_c \in \mathbb{R}^{n_u}$ is the steady state value of $u_c$. 
\\ \indent \textcolor{black}{Robust zero dynamics attacks destabilize systems' internal dynamics, leading to financial losses and significant health and safety risks. Therefore, an effective method for detecting such stealthy attacks is essential to prevent failures and performance issues in control systems. The following section presents an approach for detecting robust zero dynamics attacks in MIMO nonlinear systems.}
\section{Attack Detection \label{detection}}
\begin{figure}[h!]
	\centering
	\includegraphics[scale=0.148,trim={3.4cm 0.9cm 5.0cm 1.50cm},clip,page=1]{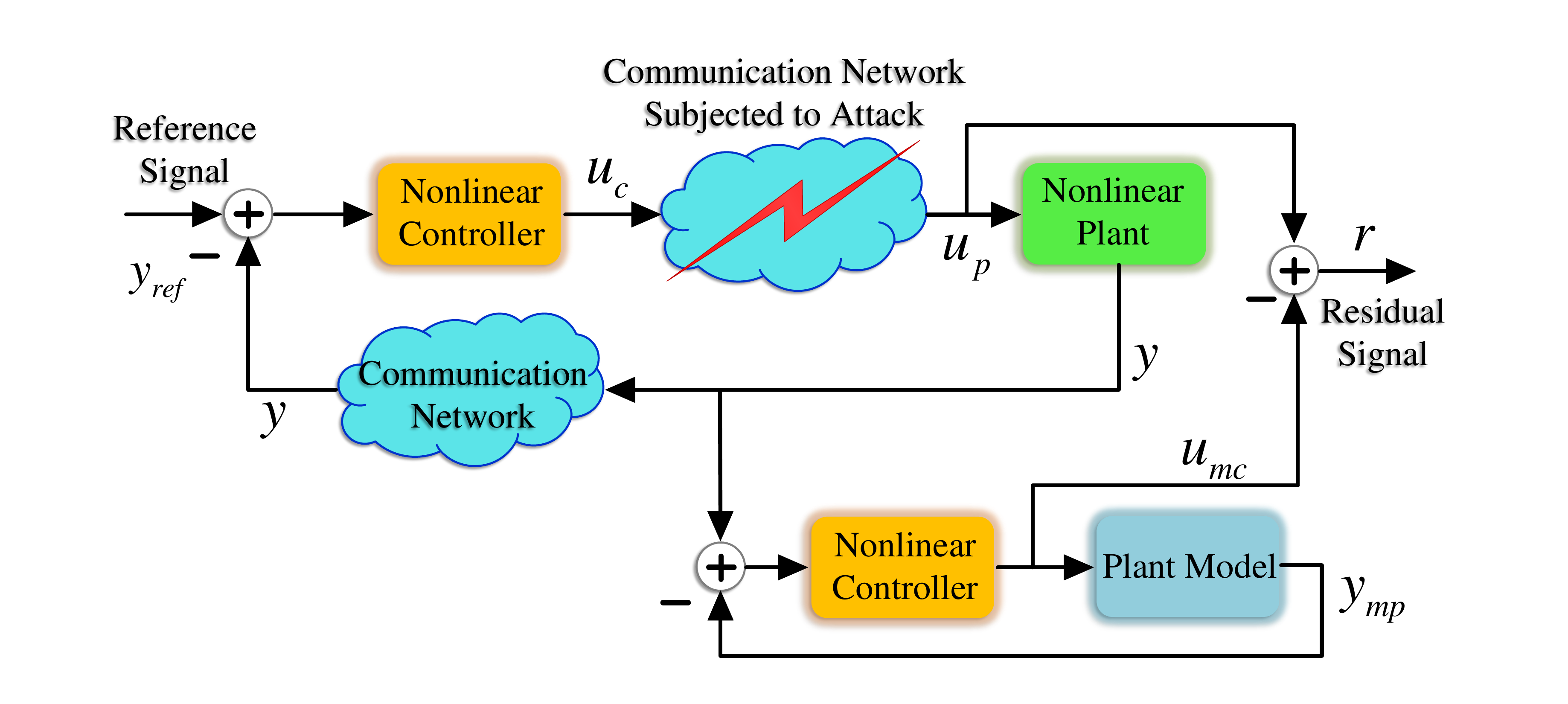}
	\caption{The general structure of the proposed robust zero dynamics attack detection strategy.}
	\label{fig::detectionstrategynonlinear}	
\end{figure}

As mentioned earlier, robust zero dynamics attacks slightly affect  systems' outputs, making the attack detection process challenging. This section presents a new method for detecting robust zero dynamics attacks without interfering with the closed-loop system. 
\\
\indent The general structure of the proposed attack detection is shown in Fig.~ \ref{fig::detectionstrategynonlinear}. For attack detection, first, the system's output is given as the input to a closed-loop system consisting of the nominal model of the system and its controller, \textcolor{black}{which can be different from the actual system and include uncertainties.} \textcolor{black}{Then, the residual signal is formed by subtracting the output of the nominal model controller from the plant input. When the system is under attack, the input signal applied to the plant includes the attack signal; therefore, it will be different from the output of the system controller. Finally, the occurrence of the robust zero dynamics attack is detected by evaluating the amplitude of the residual signal.}
\\
\indent In the following, the assumptions needed for the proposed attack detection strategy are presented.
\\
\newtheorem{assumption}{Assumption}
\begin{assumption}
	{\label{assumfirst}}
	\textcolor{black}{System and controller functions $H_{\sigma}$, $F_{\sigma}$, $G_{\sigma}$, $M$ and the nominal model functions $H_{n}$, $F_{n}$, $G_{n}$ described in (\ref{nonlinearplant}) satisfy the Lipschitz condition with  $c_5$ as an upper limit of their Lipschitz constants.}
\end{assumption}
\begin{assumption}{\label{assumsec}}
\textcolor{black}{Let's assume $q_i \in \mathbb{R}^{n_x}$ with $i \in \left\lbrace 1,2,4,6\right\rbrace$ and $q_j \in \mathbb{R}^{n_z}$ with $j \in \left\lbrace 3,5\right\rbrace$. Furthermore, consider $\mathcal{D}(\cdot)$ as the domain of a function. It is assumed that there exist constants $c_l$ for $l \in \left\lbrace 7,8,9\right\rbrace $ such that:}
\textcolor{black}{\begin{align}
	\label{assum2}
	&c_7 \parallel q_1 - q_2 \parallel \leq  \parallel G^{-1}_n(q_1) - G^{-1}_n(q_2) \parallel, \\\nonumber  &\parallel G^{-1}_n(q_1) - G^{-1}_n(q_2) \parallel \leq c_{8} \parallel q_1 - q_2 \parallel,
	\\\nonumber
	&c_9 (\parallel q_3 - q_5 \parallel  + \parallel q_4 - q_6 \parallel )	\leq  \parallel F_n(q_3,q_4) - F_n(q_5, q_6) \parallel
	\end{align}}
	with:
\textcolor{black}{\begin{align}
		q_i \in \mathcal{D}(G^{-1}_n)~~ \text{and}~~ q_j \in \mathcal{D}(F_n) 
	\end{align}}
\end{assumption}
\begin{assumption}{\label{assumthird}}
	For a nonlinear closed-loop system described in (\ref{nonlinearplant}) and (\ref{contnon}), there is a Lyapunov function $\mathcal{V}_\sigma$ and the known constant $c_4$ such that (\ref{lypassump}) holds.
\end{assumption}
\indent\begin{remark}
\textcolor{black}{ The Lipschitz condition is one of the \textcolor{black}{most widely used and least limiting assumptions} in the literature on control theory.}
	   \textcolor{black}{Note that this condition does not need to be global. \textcolor{black}{Rather, it requires the system to fulfill the Lipschitz condition within a defined interval where the state variables change.}  Therefore, if the range where the states change is known, Assumptions 1 and 2 can be simplified for a closed interval. Thus, system functions need to be Lipschitz over	
$[x_{\min}, x_{\max}]$ and $[z_{\min}, z_{\max}]$, relaxing the conditions. Here, $x_{\min}$ and $x_{\max}$ denote the minimum and maximum values of the states, and $z_{\min}$ and $z_{\max}$ express the minimum and maximum values of the system's zero dynamics in the absence of the attack, respectively. These ranges can often be identified by considering the controller's structure and estimating initial conditions. \textcolor{black}{In practice, we generally have an estimate of the range within which the system variables can vary. } Therefore, we need to evaluate the Lipschitz condition only for
those \textcolor{black}{specific} intervals. Moreover, in \textcolor{black}{ certain} cases, the desired range of the system variables is a criterion in the controller design step, which enables us to know those ranges from the beginning.} 
\end{remark}
\begin{remark}
Assumption \ref{assumsec} is the same as the bi-Lipschitz condition that functions $G^{-1}_n$ and $F_n$ need to satisfy. Since we are trying to estimate the system's input signals based on the system's output, the bi-Lipschitz condition on $G^{-1}_n$ and $F_n$ are needed. 
\textcolor{black}{The bi-Lipschitz condition is a typical condition for nonlinear functions which is widely used in the literature \cite{bilipschitz,bilipschitz1,bilipschitz2,burbano2018pinning}.}
 The existence of a Lyapunov function $\mathcal{V}_\sigma$ and constant $c_4$ satisfying (5) are needed for a robust zero dynamics attack to happen \cite{Uncertain_nonlinear}. Therefore, Assumption \ref{assumthird} only requires knowing $c_4$ or its upper bound.
\textcolor{black}{As mentioned earlier, the closed-loop system is assumed to \textcolor{black}{operate effectively } using an appropriate controller designed before, such as typical robust or adaptive methods \cite{lipschitz1,adaptive}. Therefore, one can easily find a Lyapunov function for
the system, such that \eqref{lypassump} holds.} Then, by \textcolor{black}{differentiating} the Lyapunov function and \textcolor{black}{ taking into account} the plant structure and range of uncertainties, $c_4$  or its upper bound can be \textcolor{black}{determined}. Hence, Assumption \ref{assumthird} does not introduce restrictive limitations for the problem. 
\end{remark}

\indent In the following, without loss of generality, consider the nominal model of the plant as follows:
\begin{align}
\label{mnonlinear}
\dot{z}_{mp} &= H_{n}(z_{mp},x_{mp})
\\ \nonumber
\dot{x}_{mp} &= F_{n}(z_{mp},x_{mp}) + G_n (u_{mc})\\
y_{mp} &=x_{mp} \nonumber
\end{align}
Here, $z_{mp} \in \mathbb{R}^{n_z}$ and $x_{mp} \in \mathbb{R}^{n_x}$ are state variables of the nominal model of the system. The controller output in the nominal model is defined as $u_{mc} \in \mathbb{R}^{n_u}$. Functions $H_n, F_n$, and $ G_n$ are the nominal values of the plant dynamics. 
\\\indent The controller intended to use for the nominal model of the system is considered as follows:
\begin{align}
\label{mcontnon}
\dot{s} _{mc} &= M(x_{mp}, x, s_{mc})
\\ \nonumber
u_{mc} &= K(s_{mc}, x_{mp})
\end{align}
where $s_{mc} \in \mathbb{R}^{n_c}$ stands for the vector of the controller state variables. Matrices $M$ and $K$ are as defined in (\ref{contnon}).
\textcolor{black}{The residual signal is used to detect the occurrence of the attack. Considering $u_{c} + \alpha  \in \mathbb{R}^{n_u}$ as the malicious control input of the plant, the residual signal can be defined as:} 
\begin{align}
\label{resnon}
r &\coloneqq u_{c} + \alpha - u_{mc}
\end{align}
\textcolor{black}{
\\ \indent It is worth mentioning that $u_{c} + \alpha$ is a measurable signal, because it is the input of the system which includes the controller and attack signals.}
\\\textcolor{black}{\indent In what follows, Theorem 1 \textcolor{black}{presents} the proposed detection algorithm
	based on monitoring the amplitude of the residual signal size. Theorem 1 proves
	that under \textcolor{black}{ certain } conditions, the residual signal in the absence
	of the attack converges to zero. On the other hand, when the closed-loop system is under
	attack, the residual signal may become unbounded. Hence, by
	monitoring the residual signal, the attack can be
	detected.}
\begin{theorem}
	Consider the actual system and its controller as defined in (\ref{nonlinearplant}) and (\ref{contnon}). Assume that the nominal model of the system, which can be different from the actual system, and its controller can be described as equations (\ref{mnonlinear}) and (\ref{mcontnon}). Then, if there exists constant  $c_3 > 6c_5c_4$, satisfying (\ref{lypassump}), then the residual signal defined in (\ref{resnon}) converges to zero in the absence of the attack, and converges to the attack signal in the presence of the attack.
\end{theorem}
\begin{IEEEproof}
For the proof, without loss of generality, it is assumed that the dimension of the reference signal is equal to the dimension of the system output. In case the dimension of the reference signal is smaller than the output signal, it is enough to remove the uncontrolled modes from the output and consider the resulting vector as the plant's new output.	
	\\
	\indent In summary, first, the errors between the system’s actual states and the states of the nominal model are defined. Then, because the stability of the attack free closed-loop system is previously guaranteed for normal operating conditions, conditions under which the residual signal converges to zero in the absence of attack are \textcolor{black}{derived}. Finally, it is concluded that when the robust zero dynamics attack happens, the residual signal will have a nonzero value.
		\\
	\indent In the first step, consider the attack free system, meaning that $\alpha = 0$ in (\ref{nonlinearplant}).
	\textcolor{black}{Next, let's define $\Delta_1$, $\Delta_2$, and $\Delta_3$ as follows:}
	\begin{align}
	 &\Delta_1 \coloneqq H_{n}(z_{mp}, x_{mp}) + H_{\sigma}(z - z_{mp}, x-x_{mp}) \\\nonumber &~~ ~~-H_{\sigma}(z,x)
	\\\nonumber
	&\Delta_2 \coloneqq F_{\sigma} (z, x) - F_{n}(z_{mp}, x_{mp}) - F_{\sigma} ( z - z_{mp}, x-x_{mp} )\\\nonumber &~~ ~~-G_{\sigma} (u_c - u_{mc}) + G_{\sigma}(u_c) - G_n(u_{mc})	\\\nonumber
	&\Delta_3 \coloneqq M(x,y_{\text{ref}}, s) - M(x - x_{mp}, y_{\text{ref}} - x, s- s_{mc}) \\\nonumber &- M(x_{mp}, x, s_{mc})
	\end{align}
	\indent Now, we can define the following errors:
	\begin{align}
	\label{nonlinearproof-t}
	\tilde{z} &\coloneqq z - z_{mp}, ~~~ \tilde{x} \coloneqq x - x_{mp}, ~~~	\tilde{s} \coloneqq s - s_{mc}
	\end{align}
	\indent 	\indent According to  Assumption \ref{assumfirst}, it can be concluded that 
	\begin{align}
	\parallel \Delta_1 \parallel \leq 2c_5 \parallel [\tilde{z}, \tilde{x}, \tilde{c}] \parallel \\ \nonumber
	\parallel \Delta_2 \parallel \leq 2c_5 \parallel [\tilde{z}, \tilde{x}, \tilde{c}] \parallel \\\nonumber
	\parallel \Delta_3 \parallel \leq 2c_5 \parallel [\tilde{z}, \tilde{x}, \tilde{c}] \parallel
	\\
	\end{align}
	\indent \textcolor{black}{By taking derivatives of the errors defined in (\ref{nonlinearproof-t}), the following results can be concluded:}
	\begin{align} 
	\label{nonlinearproof}
	\dot{\tilde{z}} &= \dot{z} - \dot{z}_{mp} = H_\sigma (z,x) - H_n(z_{mp}, x_{mp})  \\\nonumber &= H_{\sigma}(z - z_{mp}, x - x_{mp}) - \Delta_1
	\\\nonumber	
	\dot{\tilde{x}} &= \dot{x} - \dot{x}_{mp} \\\nonumber &= F_{\sigma}( z, x) + G_{\sigma} (u_c) - F_{n}(z_{mp}, x_{mp}) - G_n(u_{mc}) \\\nonumber&= F_{\sigma} ( z - z_{mp}, x - x_{mp}) + G_{\sigma}(u_c - u_{mc}) + \Delta_2
	\\ \nonumber
	\dot{\tilde{s}} &= \dot{s} - \dot{s}_{mc} = M(x, y_{\text{ref}}, s)-M(x_{mp}, x, s_{mc}) \\\nonumber&= M(x - x_{mp}, y_{\text{ref}} - x, s - s_{mc}) + \Delta_3
	\end{align}
	\\\indent Given the similarity between (\ref{nonlinearproof}) and the system structure in (\ref{nonlinearplant}), and according to Assumption 3, it can be concluded that $\mathcal{V}_{\sigma}$  can be considered a Lyapunov function for the system described in (\ref{nonlinearproof}). Hence, we have:
	\begin{align}
	\label{linV}
	&\parallel \dot{\mathcal{V}}_{\sigma} \parallel  \leq  \parallel
	\frac{d\mathcal{V}_{\sigma}}{d(\tilde{z}, \tilde{x}, \tilde{s})}\begin{bmatrix}
	\mathcal{S}_{11}
	\\
	\mathcal{S}_{21}
	\\
	\mathcal{S}_{31}
	\end{bmatrix}\parallel 
	\end{align}
	\textcolor{black}{with:}
	\begin{align}
	&\mathcal{S}_{11} = H_{\sigma}(z - z_{mp}, x - x_{mp}) - \Delta_1 
	\\\nonumber
	&\mathcal{S}_{21} =F_{\sigma} (x - x_{mp}, z - z_{mp}) + G_{\sigma}(u_c - u_{mc}) + \Delta_2
	\\\nonumber
	&\mathcal{S}_{31}= M(x_p - x_{mp}, y_{\text{ref}} - x) + \Delta_3	
	\end{align}
	\indent \textcolor{black}{By using Assumption \ref{assumthird}, (\ref{linV}) can be simplified as: }
	\begin{align}
\parallel \dot{\mathcal{V}}_{\sigma} \parallel  \leq -c_3 \parallel [\tilde{z},\tilde{x},\tilde{s}] \parallel ^2 + ~6c_5c_4 \parallel [z,x,s] \parallel ^2
	\end{align}
	\indent As a result, if  $c_3 > 6c_5 c_4 $, then  $\tilde{c}$ converges to zero. Therefore, the controller output of the model converges to the controller output of the actual system. Thus, the residual signal described in (\ref{resnon}) converges to zero. In case the closed-loop system is under attack, the proof would be the same because the robust zero dynamics attack has negligible effects on the output of the system. Therefore, the output of the model's controller $u_{mc}$ will converge to the output of the actual system's controller $u_c$. As a result, the residual signal defined in \eqref{resnon} will converge to the attack signal $\alpha$.
\end{IEEEproof}

 \indent \textcolor{black}{Theorem 1 proves that a robust zero dynamics attack can be detected by monitoring the amplitude of the residual signal.} \textcolor{black}{Once the robust zero dynamics attack is detected,} it is necessary to recover the system to its normal operating condition and minimize the risk of damage. The following section provides an algorithm to recover the system after detecting the attack.
\section{System Recovery \label{recovery}}
\textcolor{black}{As previously discussed, robust zero dynamics attacks destabilize a system's zero dynamics, leading to various destructive consequences. Therefore, a recovery algorithm is essential to restore the under attack system and ensure its stability. This section focuses on designing a recovery algorithm that restores the system to its normal operating condition. The recovery algorithm should simultaneously satisfy the following two conditions simultaneously:
}
\begin{enumerate}
	\item The recovery process should be such that the system output follows the desired output with an acceptable error.
	\item The recovery algorithm should stabilize \textcolor{black}{the system's zero dynamics and ensure the stability of the closed-loop system}
\end{enumerate}

\begin{figure}[h!]
	\centering
	\includegraphics[scale=0.16,trim={4.7cm 0.9cm 5.0cm 1.50cm},clip,page=1]{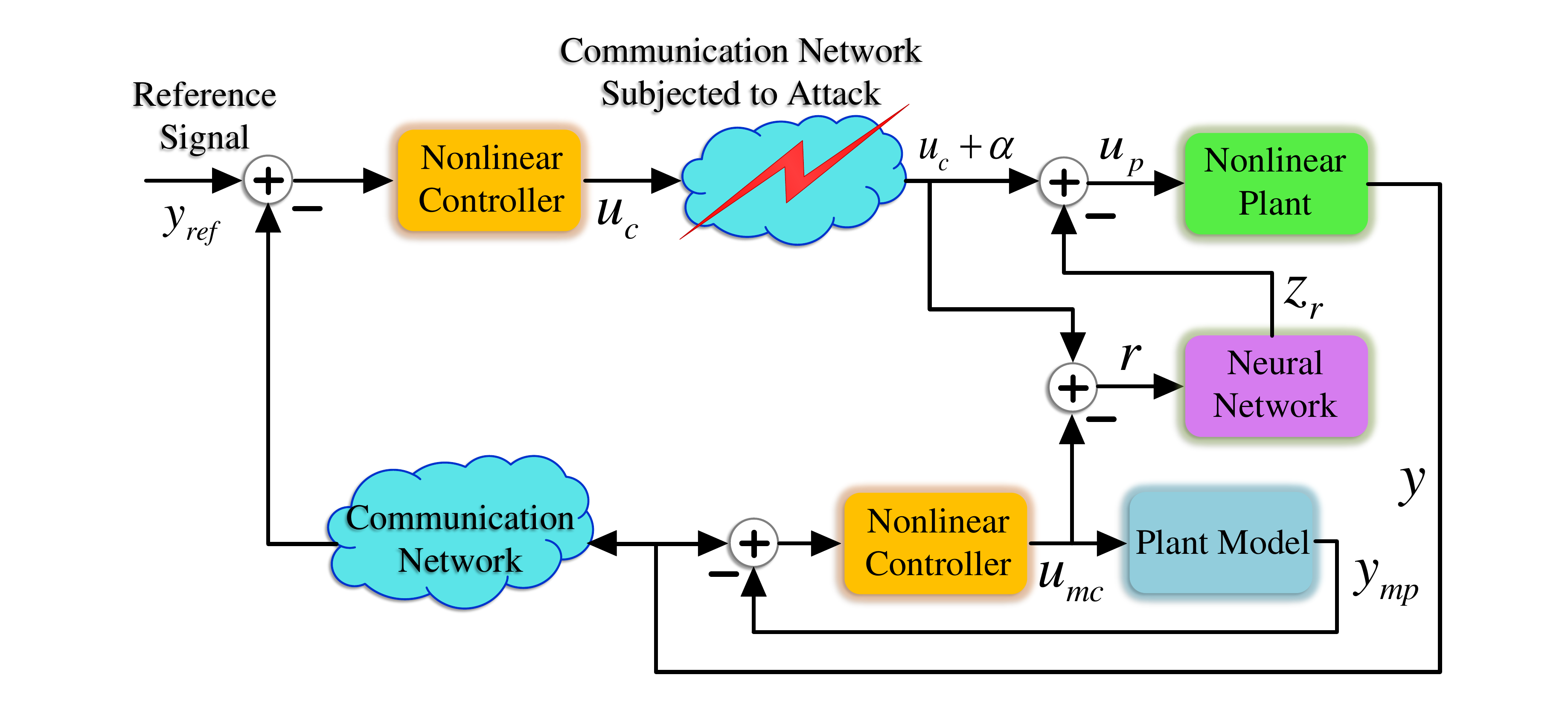}
	\caption{ The general structure of the proposed recovery algorithm.}
	\label{fig::recoverystrategynonlinear}	
\end{figure}

\indent A general scheme of the proposed method for the system recovery and mitigation of the destructive effects of the robust zero dynamics attacks is shown in Fig.~\ref{fig::recoverystrategynonlinear}. Once the occurrence of the attack is detected, the attack signal is estimated using the system states as the input for a neural network model, which can approximate the attack signal. Moreover, the residual signal is utilized to update the neural network weights with one hidden layer without need for any prior data. Finally, the approximated robust zero dynamics attack signal obtained from the neural network is subtracted from the system input to recover the closed-loop system.

 \indent \textcolor{black}{ A three-layer neural network composed of the input, hidden, and output layers is used to estimate the attack signal. The output of the neural network is the recovery signal, which can be considered as an approximation of the attack signal. The recovery signal is subtracted from the system input to mitigate the effects of the attack. States of the system are given as the input to the neural network, while the residual signal is used to update the weights adaptively. The hidden layer is responsible for processing the input using weights of the neural network and some nonlinear functions, named activation functions. Weights of the neural network are updated using online adaptation rules.
 }  
\\
\indent \textcolor{black}{Let's consider $\lambda$ and $e$ as the learning rate and a measurable error signal, which are obtained from the recovery and residual signals, used to update the parameters of the neural network. Then, the weights of the neural network are updated by the following rule:}
\textcolor{black}{\begin{align}
\label{updateart}
&\nonumber \dot{w} = \lambda  \frac{\partial G^{-1}_n (F_n(w, x^*))}{\partial w} \frac{1}{\parallel \frac{\partial G^{-1}_n (F_n(w, x^*))}{\partial w } \parallel ^2} e \\\nonumber &+ H_n(w+z^*,x^*)
\\
& e  \coloneqq \alpha - G^{-1}_{n}(F_n(w,x^*)) + u_c - u_{mc} 
\end{align}}
\indent \textcolor{black}{
 In the hidden layer, $F_n(\cdot)$ is considered as an activation
 function. As a result, information of the nominal model
 can be used to compute the output of the hidden
 layer of the neural network. The output of the hidden layer can be
 computed by giving states and weights of the neural network as the input
 to the activation functions. Hence, the output of the hidden
 layer is $F_n(w,x^*)$, in which $\omega$ stands for the weights of the neural network
 described in (\ref{updateart}).\\
  \indent As mentioned before, the output of the neural network is considered as the recovery signal. In the following, a nonlinear function is needed to be applied to the output of the hidden layer output to compute the recovery signal or the output of the neural network model. In the output layer, $G_n^{-1}(\cdot)$  is considered as the nonlinear function, which applies to the
 hidden layer's output. Please note that $G_n^{-1}(\cdot)$ is part of the nominal model dynamics and is known. Hence, the output of the neural network model, i.e., the recovery signal, can be obtained as: 
\begin{align}
	\label{zr}
z_r \coloneqq G_n^{-1}(F_n(w, x^*))
\end{align} }
\\
\indent \textcolor{black}{ To recover the system, the output of the neural network is subtracted from the plant’s input. Therefore, in the presence of the robust zero dynamics attack, the following input is given to the system:
\begin{align}
\label{upattack}
u_p = \alpha - G^{-1}_{n}(F_n(w, x^*)) + u_c
\end{align}}
\textcolor{black}{\indent It must be noted that the main difference between the input $u_P$ defined in \eqref{upattack} and the input signal defined in the previous section is the recovery signal $z_r$. The input of the nominal model in the presence of the attack is the same as the output of the nominal model controller described in (\ref{mcontnon}). We have the output of the neural network model. Hence, error signal $e$ described in (\ref{updateart}) can be obtained by subtracting the input of the nominal model input from the actual system input. Since the nominal model of the system is available, $ \frac{\partial G^{-1}_n F_n(w, x^*)}{\partial w}$ is a known term and can be used to update the weights of the neural network.}\\
\indent \textcolor{black}{
Now, let’s consider previously defined variables $z$, $x$,  $F_{\sigma}$,  $G_{\sigma}$, $y$, $u_c$,  and $\alpha$. Assuming $z_r$ as the output of the neural network described in (\ref{zr}), the closed-loop system under the recovery algorithm can be described as
follows:
}
\begin{align}
\label{attack+recovery}
\dot{z} &= H_{\sigma}(z,x)
\\\nonumber
\dot{x} &= F_{\sigma}(z,x) + G_{\sigma} (u_c - z_r + \alpha) \\\nonumber
y &= x\nonumber
\end{align} 
\indent  Theorem 2 proves the effectiveness of the proposed recovery algorithm as well as the stability of the closed-loop system. It is shown that under specific conditions, the zero dynamics of the system subjected to the stealthy attack remain bounded, and the output of the system follows the reference signal with an acceptable performance.
\begin{theorem}
Consider the closed-loop system described in (\ref{nonlinearplant}). If Assumptions \ref{assumfirst}-\ref{assumthird} and  $c_3 \geq c_4c_5(3+ c_5 c_8) + \lambda$, $\lambda \leq \frac{c_4}{c_7c_9}$ hold, \textcolor{black}{then the neural network model defined in (22) estimates the robust zero dynamics attack signal with a limited error (i.e., $\parallel z_r - \alpha \parallel $ remains bounded).} \textcolor{black}{ Moreover, the zero dynamics of the recovered system remains bounded, the output of the
recovered system follows the reference signal, and the closed-loop system stays stable.}
\end{theorem}
\begin{IEEEproof}
Let us use the Lyapunov stability approach to
study the boundedness of errors in the presence of the attack. In the first step, by adding neural network recovery terms to the control signal in (\ref{attack+recovery}), one has:
	\begin{align}
	\label{recovery_attack_plant}
	 &\dot{x} = F_{\sigma}(z,x)  + G_{\sigma} (u_c + \alpha - z_r)
	\\\nonumber
	&\dot{x} =   F_{\sigma}(z',x) + G_{\sigma}(u_c) + \Delta_4
	\\\nonumber
	&z' \coloneqq z + \tilde{\delta} - w
		\\\nonumber
	&\dot{z}' = \dot{z} + \dot{\tilde{\delta}} - \dot{w}
	\nonumber 
	\end{align}
	\textcolor{black}{where 	$\Delta_4$ is considered as:}
	\begin{align}
	\Delta_4 \coloneqq - F_{\sigma}(z',x) +  G_{\sigma} (u_c + \alpha - z_r) - G_{\sigma}(u_c)
	\end{align}
	\indent \textcolor{black}{
	By replacing $\dot{\tilde{\delta}}$ and $\dot{w}$ from (\ref{nonlinearattack}) and (\ref{updateart}), $\dot{z}'$ in (\ref{recovery_attack_plant}) can be simplified as follows:
	}
	\begin{align}
	\label{temp_dot_z}
   &\dot{z}' = H_{\sigma}(z,x) - \mathcal{G} + \mathcal{T} 
   \end{align}  
   with:
   \begin{align}
   &\mathcal{T} =  H_n(\tilde{\delta} + z^*_n, x^*) - H_n(w+z^*_n, x^*)\\\nonumber
    &\mathcal{G} = \lambda  \frac{\partial G^{-1}_n F_n(w, x^*)}{\partial w} \frac{1}{\parallel \frac{\partial G^{-1}_n F_n(w, x^*)}{\partial w} \parallel ^2} e \\\nonumber
	\end{align}
	\indent{\textcolor{black}{After some manipulations, (\ref{temp_dot_z}) can be written as: }}
	\begin{align}
	&\dot{z}' = H_{\sigma}(z',x) - \mathcal{G}+ \Delta_5 
\end{align}
\textcolor{black}{where $\Delta_5$ is defined as:}
\begin{align}
	&\Delta_5 \coloneqq - H_{\sigma}(z',x) +  H_{\sigma}(z,x) + \mathcal{T} 
	\end{align}
	\indent  According to Assumption \ref{assumfirst}, it can be concluded that
	$\parallel \Delta_4 \parallel  \leq  \parallel (c_5 + c^2_5 c_8)[s, x, z']\parallel $ and $ \parallel \Delta_5 \parallel  \leq  \parallel 2c_5[s, x, z']\parallel $.
	Moreover, using Assumption \ref{assumthird}, for a closed-loop system in
	the absence of robust zero dynamics attacks, there is a Lyapunov function $\mathcal{V}_{\sigma}$
	satisfying (\ref{lypassump}).
	\\\indent\textcolor{black}{  
	 Now, we can consider the following Lyapunov function candidate to prove the stability of the closed-loop system after applying the recovery algorithm:}

	\begin{align}
	\label{lypnon}
	\mathcal{V}' = \mathcal{V}_{\sigma}(x, z', s)+ (\alpha - z_r)^T(\alpha - z_r) + \mathcal{L}^T\mathcal{L}
	\end{align}
	Here $\mathcal{L}$ can be defined as:
	\begin{align}
&\dot{\mathcal{L}} = -\frac{sign(\mathcal{L})}{\parallel \mathcal{L} + \varepsilon \parallel}  \parallel r - z_r \parallel (\frac{ c_{8} c_5  }{c_7c_9} ( \parallel z_r \parallel + \parallel r \parallel + \mathcal{K}))&
\\\nonumber
&\mathcal{K}=-c_7 \parallel F_n(z^*, x^*) \parallel+ \frac{2c_7c_9}{c_5} \parallel H_n(w + z^*_n,x^*) \parallel + \\\nonumber&  \;\;\;\;\;\;\;\;c_7c_9 \parallel z^* \parallel
		\end{align}
with $\varepsilon$ as a positive small number to avoid division by zero.
	We use the auxiliary variable $\mathcal{L}$ in the Lyapunov function to simplify the proof.
	Now, by taking the time derivative of (\ref{lypnon}) we have:
	\begin{align}
	\label{Vbe}
	\dot{\mathcal{V}}' = \dot{\mathcal{V}}_{\sigma}(x, z', s)+ 2(\alpha - z_r)^T \frac{d}{dt} ( \alpha - z_r) + 2\mathcal{L}^T \dot{\mathcal{L}}
	\end{align}
	\indent In the following, the terms appeared in the Lyapunov derivation are examined and simplified separately.
	The goal of this step is first to find upper bounds for the different terms that appeared in the derivative of the Lyapunov function. 
	 First, consider $\dot{\mathcal{V}}_{\sigma}(x, z', s)$. According to the similar structure of (\ref{recovery_attack_plant}) with the attack free closed-loop system \textcolor{black}{ (\ref{nonlinearplant}) and (\ref{contnon})}, and under Assumptions \ref{assumfirst}-\ref{assumthird}, (\ref{Vbe}) can be simplified as:
	\begin{align}
	\label{dlypnon}
	\parallel \dot{\mathcal{V}}_{\sigma} \parallel &\leq \parallel \frac{d\mathcal{V}}{d(z',x, s)}\begin{bmatrix}
	\mathcal{V}_{1}
	\\
	\mathcal{V}_{2}
	\\
	\mathcal{V}_{3}
	\end{bmatrix} \parallel \\ \nonumber & \leq (-c_3 + 3c_5c_4  + c^2_5 c_4 c_8 + \parallel \mathcal{G} \parallel c_4)   \parallel [x, s, z'] \parallel ^2  
		\end{align} 
	with $\mathcal{V}_1$, $\mathcal{V}_2$, and $\mathcal{V}_3$ defined as:
	\begin{align}
	\mathcal{V}_{1} &= H_{\sigma}(z',x) + \Delta_5 - \mathcal{G}
	\\\nonumber
	\mathcal{V}_{2} &=  F_{\sigma}(z',x) + G_{\sigma}(u_c) + \Delta_4
	\\\nonumber
	\mathcal{V}_{3}  &= M(x_p - x_{mp}, y_{\text{ref}} - x) + \Delta_3	
	\end{align}
	\indent \textcolor{black}{In what follows, after completing the square for $c_4  \parallel \mathcal{G} \parallel  \parallel [x, s, z'] \parallel ^2  $, one has:}
\begin{align}
c_4  \parallel \mathcal{G} \parallel   \parallel [ x, s, z'] \parallel ^2 &=\nonumber \mathcal{H} ~ [-(\parallel e \parallel + \parallel [x, s, z'] \parallel ) ^2\\ &\;+ \parallel e \parallel^2 +  \parallel [x, s, z'] \parallel ^2 ]
\\\nonumber
\mathcal{H} &= \frac{\lambda c_4 } {\parallel {\frac{\partial G^{-1}_n F_n(w, x^*)}{\partial w} \parallel} } \nonumber
\end{align} 
	\indent In the next step, we will simplify terms $(\alpha - z_r)$ and $\mathcal{L}^T \dot{\mathcal{L}}$ in (\ref{Vbe}). After taking partial derivatives of $(\alpha - z_r)$ in (\ref{Vbe}) and the proposed adaptation law in (\ref{updateart}), we have:
	\begin{align}
	\label{a-gn}
	 \frac{d}{dt}(\alpha - z_r)&=  -\mathcal{I}_w (H_n(w + z^*_n,x^*) + \mathcal{G})  \\\nonumber &\;+ \mathcal{I}_\delta H_n(\tilde{\delta} + z^*, x^*)
	\end{align}
	where:
	\begin{align}
	&\mathcal{I}_w = \frac{\partial G^{-1}_n F_n(w, x^*)}{\partial w}
	\\\nonumber
	&\mathcal{I}_\delta = \frac{\partial G^{-1}_n F_n(\tilde{\delta}, x^*)}{\partial \tilde{\delta}}
	\end{align}
	Therefore:
	\begin{align}
	\label{tempeq}
     (\alpha - z_r)^T \frac{d}{dt}(\alpha - z_r)&= (\alpha - z_r)^T(  \mathcal{I}_\delta H_n(\tilde{\delta} + z^*,x^*)\\\nonumber &-\mathcal{I}_w (H_n(w + z^*_n,x^*) + \mathcal{G}))
	\end{align}
	
	\indent To simplify $(\alpha - z_r)^T$ $\frac{d}{dt}(\alpha - z_r)$, the terms that appeared in (\ref{tempeq}) will be analyzed. Let us consider terms in (\ref{tempeq}) containing error signal $e$. As a result, the following can be concluded:
	\begin{align}
	\label{c6-t}
 & -\parallel (\alpha - z_r)^T ( ~ \mathcal{I}_w ~\mathcal{G}) \parallel  = \\\nonumber& -\lambda \parallel  (\alpha -z_r)^T (\alpha -z_r)  +   (\alpha-z_r)^T( u_c- u_{mc}) \parallel 
	\end{align}
	\indent \textcolor{black}{Again, after completing the square for (\ref{c6-t}), we can write:}
	\begin{align}
	\label{c6}
&-\parallel (\alpha - z_r)^T ( ~ \mathcal{I}_w ~ \mathcal{G}) \parallel  \leq   - \frac{\lambda}{2} \parallel  (\alpha - z_r) \parallel^2  \\\nonumber
&- \frac{\lambda}{2} \parallel (\alpha - z_r + u_c -u_{mc}) \parallel ^2 +  \frac{\lambda}{2}\parallel (u_c - u_{mc})\parallel ^2  
	\end{align}	
	\indent Now, consider the error free terms in (\ref{tempeq}):
	\begin{align}
	\label{L}
	\nonumber&\parallel- \mathcal{I}_w H_n(w + z^*_n,x^*) + \mathcal{I}_\delta H_n(\tilde{\delta} + z^*, x^*) \parallel= \\\nonumber & \parallel \mathcal{I}_\delta (H_n(\tilde{\delta} + z^*, x^*) -  H_n(w + z^*_n,x^*))	\\\nonumber &+ (\mathcal{I}_\delta -\mathcal{I}_w)  H_n(w + z^*_n,x^*) \parallel \\ &\leq c_{8} c_5 \parallel \tilde{\delta} - w \parallel  + ~ 2c_{8} H_n(w + z^*_n,x^*)
	\end{align}
	\indent It is worth mentioning that there is no \textcolor{black}{known} upper bound on $\parallel \tilde{\delta} - w \parallel$. Therefore, according to Assumption 2, an upper bound for $\parallel \tilde{\delta} - w \parallel$ can be found as follows:
	\begin{align}
	\nonumber&\parallel z_r +  \alpha \parallel \geq c_7 \mathcal{F}_n(w, \tilde{\delta}) + c_7 \parallel F_n(z^*, x^*) \parallel \geq \\\nonumber &c_7c_9 \parallel w -\tilde{\delta} - z^* \parallel  + ~ c_7 \parallel F_n(z^*, x^*) \parallel \geq \\\nonumber & c_7c_9 (\parallel w -\tilde{\delta} \parallel - \parallel z^* \parallel) +  c_7 \parallel F_n(z^*, x^*) \parallel 
\end{align}
with:
\begin{align}
	&\mathcal{F}_n(w, \tilde{\delta}) = \parallel F_n (w, x^*) - F_n(\tilde{\delta} + z^{*}_n , x^*) \parallel 
	\end{align}
	Therefore, we can obtain:
	\begin{align}
	\label{temp}
	& \parallel w - \tilde{\delta} \parallel \leq \frac{1}{c_7c_9}  (- c_7 \parallel F_n(z^*, x^*) \parallel \\\nonumber  & + \parallel G^{-1}_n(F_n(w,x^*)) \parallel + \parallel G_n^{-1}(F_n(\tilde{\delta} + z^{*}_n , x^*) \\\nonumber &- F_n(z^*, x^*))  \parallel) + \parallel z^* \parallel
	\end{align}
	\indent Since $G_n^{-1}(F_n(\tilde{\delta} + z^{*}_n , x^*) - F_n(z^*, x^*)) $ in (\ref{temp}) is not available, the estimation of the attack signal $G^{-1}_n(F_n(w,x^*))$ is used to find an upper bound for $\parallel w - \tilde{\delta} \parallel$:
	\begin{align}
	\label{temp2}
	& \parallel w - \tilde{\delta} \parallel \leq \frac{1}{c_7c_9} (\parallel G^{-1}_n(F_n(w,x^*)) \parallel \\\nonumber &+ \parallel r \parallel - c_7 \parallel F_n(z^*, x^*) \parallel) + \parallel z^* \parallel
	\end{align}
	\\\indent As we saw in Theorem 1, the residual signal converges to the attack signal and we will have: 
	\begin{align}
	&\mathcal{L}^T\dot{\mathcal{L}} + \parallel (\alpha - z_r)^T(- \mathcal{I}_w H_n(w + z^*_n,x^*)\\\nonumber &+  \mathcal{I}_\delta H_n(\tilde{\delta} + z^*, x^*) )\parallel \leq 0
	\end{align}
	\indent Now, according to (\ref{dlypnon}), (\ref{a-gn}) and (\ref{L}), it can be concluded that if $c_3 $ and $\lambda $ are selected in a way that $c_3 \geq c_4c_5(3+ c_5 c_8) + \lambda$ , $\lambda \leq \frac{c_4}{c_7c_9}$, then the derivative of the Lyapunov function is negative. Therefore, the estimation error $\parallel \alpha -  z_r \parallel$ remains bounded.
\end{IEEEproof}

\indent \textcolor{black}{The neural network used for the recovery process works in parallel with the closed-loop system without interfering or impacting system controller and its performance. Moreover, the process of updating the neural network weights does not need an accurate model of the system, making it robust against model uncertainties and environmental imperfections. We just need to know a nominal model of the system, which can be different from the actual system and have parametric or structural uncertainties.}
\begin{remark}
 \textcolor{black}{As mentioned before, the residual signal converges to the attack signal under ideal conditions. However, because of potential imperfections and disturbances, the residual signal may differ from the attack signal. In this regard, the neural network and the nominal model of the system enable us to have a better estimation of the attack signal. Then the estimated signal is used to recover the closed-loop system.}
\end{remark}
\begin{remark}
 \textcolor{black}{Conditions stated for the parameters used in Theorems 1 and 2 can be easily satisfied after designing an appropriate closed-loop controller and choosing a proper Lyapanuv function. In fact, since the proposed strategy is applicable to systems with previously designed controllers, irrespective of the control method, the Lyapunov function used for the control system design can be used to to satisfy the conditions introduced in Theorems 1 and 2.}
\end{remark}
\section{Simulation Results \label{simulation}}
In this section, a four-tank system is considered to simulate the proposed method and show its effectiveness. The selected system is a MIMO nonlinear system, which can be described as follows:
\begin{align}
\label{nonsys}
&\dot{z_{\sigma}} = H_{\sigma}(z_{\sigma},x)
\\ \nonumber
&\dot{x} = F_{\sigma}(z_{\sigma},x) + G_{\sigma}(u_p)
\\\nonumber
&y = x
\end{align}
\textcolor{black}{Here \textcolor{black}{$u_p \in \mathbb{R}^2 $ and $\alpha \in \mathbb{R}^2 $ are the control input (input voltage of the pumps) and the attack signal \cite{Uncertain_nonlinear}. $x \in \mathbb{R}^2 $ and $z_\sigma \in \mathbb{R}^2 $ stand for the output and zero dynamics of the system, respectively. State variables of the system represent the water level of tanks. Functions $H_{\sigma}, F_{\sigma}$, and $G_{\sigma}$ in (\ref{nonsys}) are defined as:}
\begin{align}
\label{nonsys2}
	&H_{\sigma}(z_{\sigma},x) = \begin{bmatrix}
	H_{\sigma _1}(z_{\sigma},x) \\
	H_{\sigma _2}(z_{\sigma},x)
	\end{bmatrix} 
	\\\nonumber 
	&H_{\sigma _1}(z_{\sigma},x)  = -\frac{\sqrt{2g}a_L}{A_L} \sqrt{z_{\sigma,1} + T_{\sigma ,2} x_2}\\\nonumber &+ \frac{\sqrt{2}g(1 - \sigma_2)a_R}{\sigma_2 A_L}(\sqrt{x_2} - \sqrt{z_{\sigma,2} + T_{\sigma,1}x_1})
	\\\nonumber
	&H_{\sigma _2}(z_{\sigma},x) = 	-\frac{\sqrt{2g} a_R}{A_R} \sqrt{z_{\sigma,2} + T_{\sigma , 1} x_1} \\\nonumber &+ \frac{\sqrt{2g}(1 - \sigma_1)a_L}{\sigma_1 A_R} (\sqrt{x_1} - \sqrt{z_{\sigma ,1} + T_{\sigma , 2}x_2})
	\nonumber
\end{align}
and
\begin{align}
\label{nonsys22}
	&F_{\sigma}(z_{\sigma},x) = \begin{bmatrix}
	F_{\sigma  1}(z_{\sigma},x)
	\\
	F_{\sigma  2}(z_{\sigma},x)
	\end{bmatrix} 
	\\\nonumber 
	&F_{\sigma  1}(z_{\sigma},x) = \frac{\sqrt{2g}a_L}{A_L} (-\sqrt{x_1} + \sqrt{z_{\sigma , 1} + T_{\sigma , 2} x_2})
	\\\nonumber
	&F_{\sigma 2}(z_{\sigma},x) = \frac{\sqrt{2g}a_R}{A_R}(-\sqrt{x_2} + \sqrt{z_{\sigma ,2} + T_{\sigma ,1}x_1})
	\\\nonumber
	&G_{\sigma}(u_p) = diag(g_{\sigma 1}, g_{\sigma  2})u_p = diag(\frac{\sigma_1 k_1}{A_L}, \frac{\sigma_2 k_2}{A_R})u_p
\end{align}
with $(A_l, A_R) = (28,32)cm^2$, $(a_l,a_R) = (0.071,0.057)cm^2$, $(k_1,k_2) = (3.14,3.29) \frac{cm^3}{Vs}$, $(\sigma_1, \sigma_2) = (0.43, 0.34)$, and $g = 9.81$. }
 To model uncertainties, the parameters of the nominal model are considered as follows:
\begin{align}
(A_L, A_R) &= (30, 34),\\
(a_L, a_R) &= (0.101, 0.057),\nonumber(k_1, k_2) &= (3.14, 3.5)
\end{align} 
\\\indent A PI controller is considered for the system and its nominal model with the following structure 	\cite{Uncertain_nonlinear}:
\begin{align}
\label{contsim}
&\dot{c} = y -r
\\\nonumber
&u_c = \begin{bmatrix}
\sigma_1 k_1 & (1 - \sigma_2) k_2 \\
(1 - \sigma_1)k_1 & \sigma_2 k_2
\end{bmatrix} ^{-1} \begin{bmatrix}
a_L \sqrt{2gr_1}
\\
a_R \sqrt{2gr_2}
\end{bmatrix} \\\nonumber &~~~~+ k_p (y - r) + k_i c
\\ \nonumber
&k_p = diag(0.75, -0.06),\\ \nonumber
&k_i = diag(0.0068 , -0.00027) ,\\ \nonumber
&y = \begin{bmatrix}
x_1 \\ x_2
\end{bmatrix}
\end{align}
\\\indent For simulation, the output of the system is given as a 
reference signal to the nominal model which includes the controller (the PI controller introduced above), as shown in Fig.~\ref{fig::detectionstrategynonlinear}. \textcolor{black}{The objective of the system controller is enabling the system to track a reference signal with a bounded error while all state variables remain above zero.
The dynamics of the four-tank system defined in (\ref{nonsys}) and (\ref{nonsys2}) include constant values and radical functions. It can easily be shown that these terms satisfy both Lipschitz and bi-Lipschitz conditions (Assumptions \ref{assumfirst} and \ref{assumsec}) over any closed interval greater than zero. Hence, the four-tank system satisfies Assumptions \ref{assumfirst} and \ref{assumsec}. Moreover, according to \cite{robust11}, it is proven that the four-tank system satisfies Assumption \ref{assumthird} as well.} In simulations, the closed-loop system is assumed to be the under attack signal described in \eqref{nonlinearattack} \textcolor{black}{ from $t = 700$s to $t = 1000$s.}
\\\indent The following indices are provided to evaluate the performance of the proposed method:
\begin{enumerate}
	\item  Attack detection time: it is defined as the difference between the time instant when the residual signal is nonzero and the time instant at which the attack happens.
	\item \textcolor{black}{Attack success rate: this index shows the effectiveness of the attack by using the system's zero dynamics in different scenarios. This index is defined as:}
	\textcolor{black}{\begin{eqnarray}
		\gamma \coloneqq \frac{\parallel z_r - z_n \parallel_\infty}{\parallel z_a - z_ n \parallel_\infty}
	\end{eqnarray}}
	\textcolor{black}{\noindent with $z_r \in \mathbb{R}^{n_z}$ and $z_a \in \mathbb{R} ^{n_z}$ are the zero dynamics of the system subjected to the attack under the recovery algorithm and without the recovery algorithm, respectively. $z_n \in \mathbb{R}^{n_z}$ is considered as the zero dynamics of the system in normal operating condition without the attack.}
\end{enumerate}

\indent The first indicator represents attack detection time, while the second one shows the effectiveness of the proposed recovery algorithm. The closer these indicators are to zero, the better the algorithm's performance. \textcolor{black}{In subsection \ref{const}, the performance of the proposed strategy for a constant reference signal is presented. Subsection \ref{noise} shows the algorithm's robustness against noise disturbances. Subsection \ref{timevar} evaluates the effectiveness of the proposed strategy for a time-varying reference signal. Subsection \ref{compare} compares the efficiency of the proposed method with the resilient method presented in \cite{simulation-compare}.}
\subsection{Performance Evaluation with a Constant Reference \label{const}}
\begin{figure}[h!]
	\centering
	\includegraphics[width=1\linewidth]{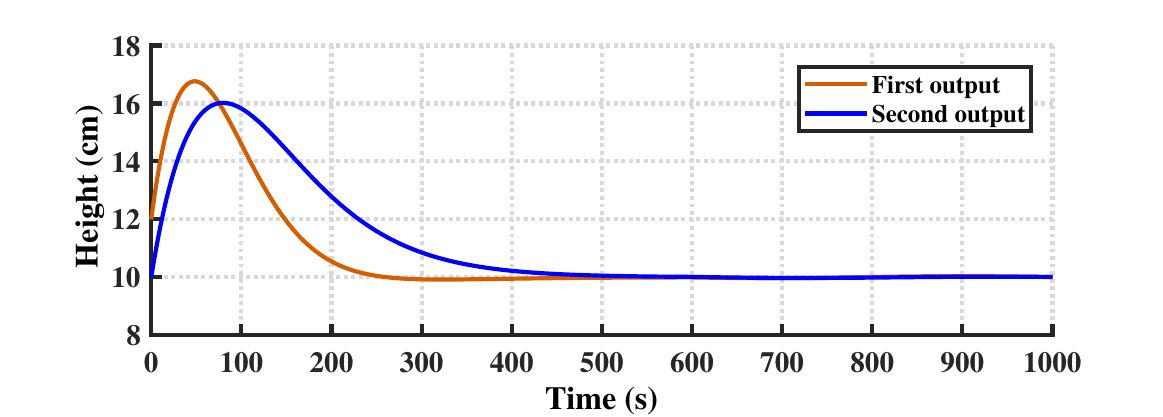}
	\caption{\textcolor{black}{System output in scenario A: The output of the closed-loop system with a constant reference.}}
	\label{fig::outm2}	
\end{figure}

\begin{figure}[h!]
	\centering
	\includegraphics[width=1\linewidth]{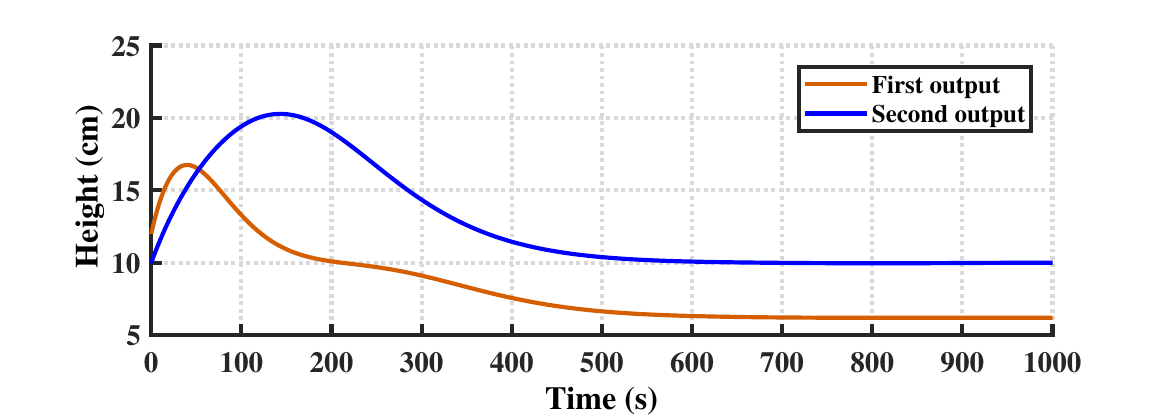}
	\caption{\textcolor{black}{Model output in scenario A: The output of the model with a constant reference.}}
	\label{fig::outm1}	
\end{figure}

 \textcolor{black}{In this subsection, a constant value of 10 is considered for the closed-loop system's reference signal. The outputs of the system and the nominal model are depicted in Fig.~(\ref{fig::outm2}) and  (\ref{fig::outm1}). It can be seen that the model output follows the system output with a limited error. This error is because of the uncertainties and differences between the nominal model and the actual system. An adversary inject a zero dynamics attack to the system at $t = 700$s to affect the performance of the closed-loop system (making tanks empty in a four-tank system). The simulation results show that the output tracks the reference signal in the presence of the attack, which verifies the effectiveness of the recovery algorithm.} Hence, the recovery algorithm does not significantly affect the system’s steady-state behavior. \textcolor{black}{As a result, the proposed strategy can be applied to a closed-loop system in parallel with the plant's controller to keep the system stable and mitigate the destructive effects of the attack without interfering with the closed-loop system performance.} 
 
\begin{figure}[h!]
	\centering
	\includegraphics[width=1\linewidth]{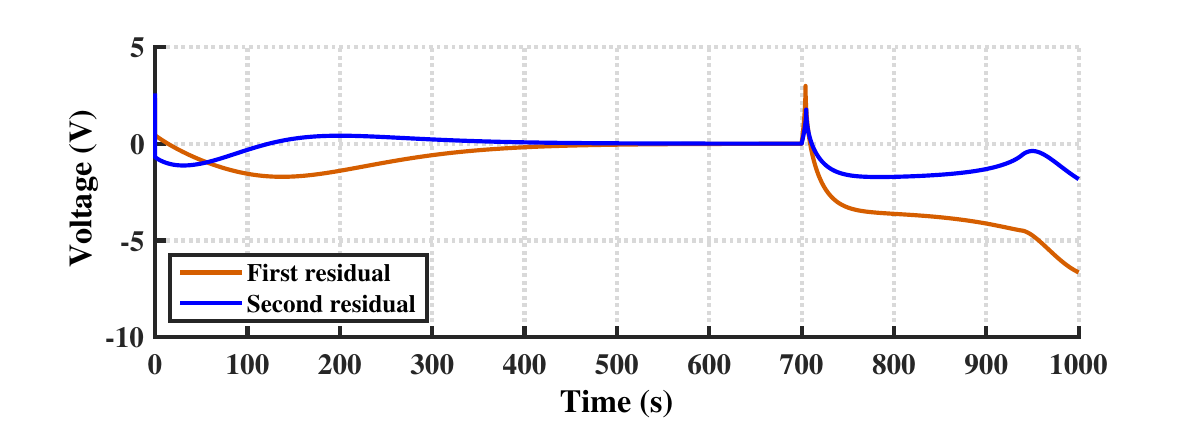}
	\caption{\textcolor{black}{Residuals in scenario A: Residual signals used to detect the occurrence of the attack and update the neural network weights.}}
	\label{fig::attacknon}	
\end{figure}

\begin{figure}[h!]
	\centering
	\includegraphics [width=1\linewidth]{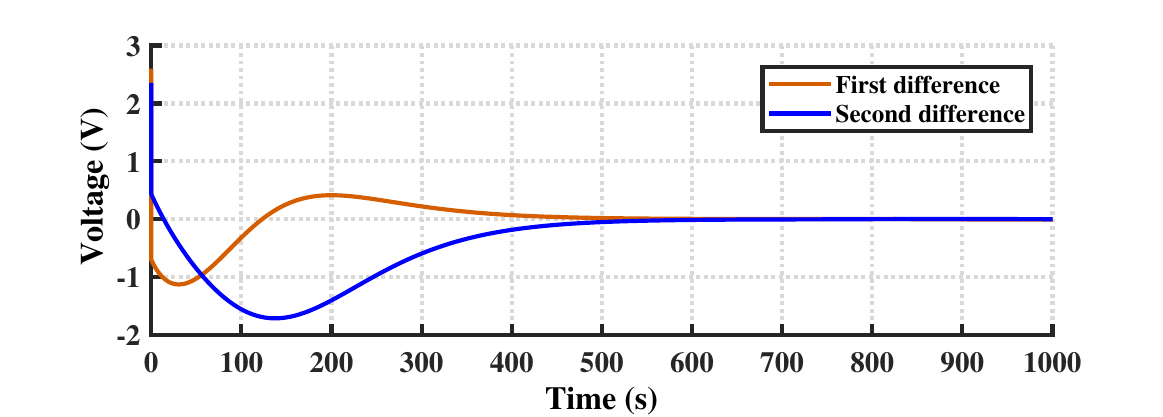}
	\caption{\textcolor{black}{Difference between residuals and attack signals in scenario A: It can be seen that after a short transients, residuals converge to attack signals.}}
	\label{fig::resnon}	
\end{figure}

\begin{figure}[H]
	\centering
	\includegraphics[width=1\linewidth]{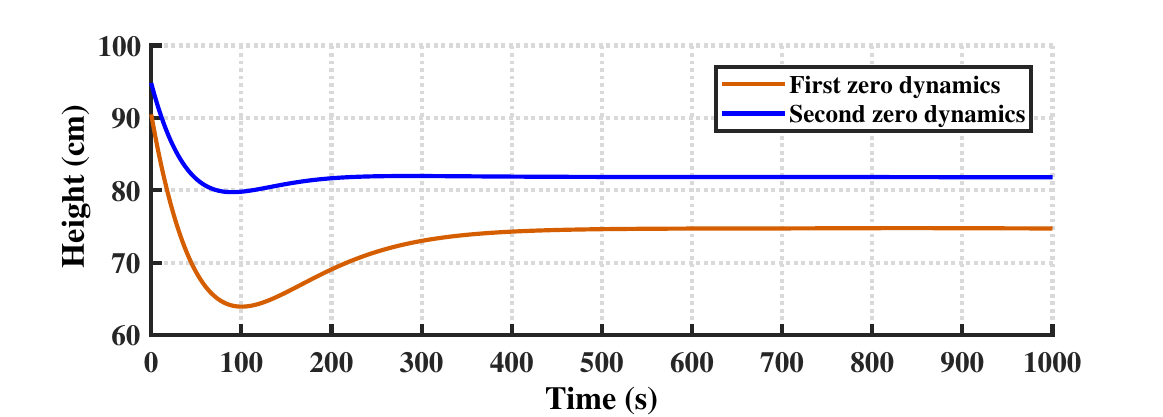}
	\caption{\textcolor{black}{Zero dynamics the system in scenario A: Zero dynamics of the system stay stable under the proposed recovery algorithm.}}
	\label{fig::z1}	
\end{figure}

\begin{figure}[h!]
	\centering
	\includegraphics[width=1\linewidth]{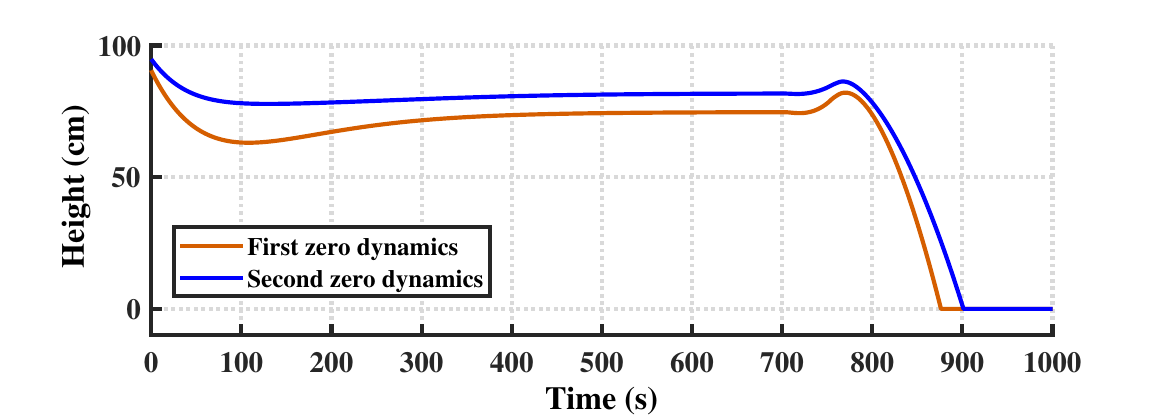}
	\caption{\textcolor{black}{Zero dynamics of the system in scenario A: Zero dynamics of the system show inappropriate behavior in the case without the recovery algorithm.}}
	\label{fig::z2}	
\end{figure}

\begin{figure}[h!]
	\centering
	\includegraphics[width=1\linewidth]{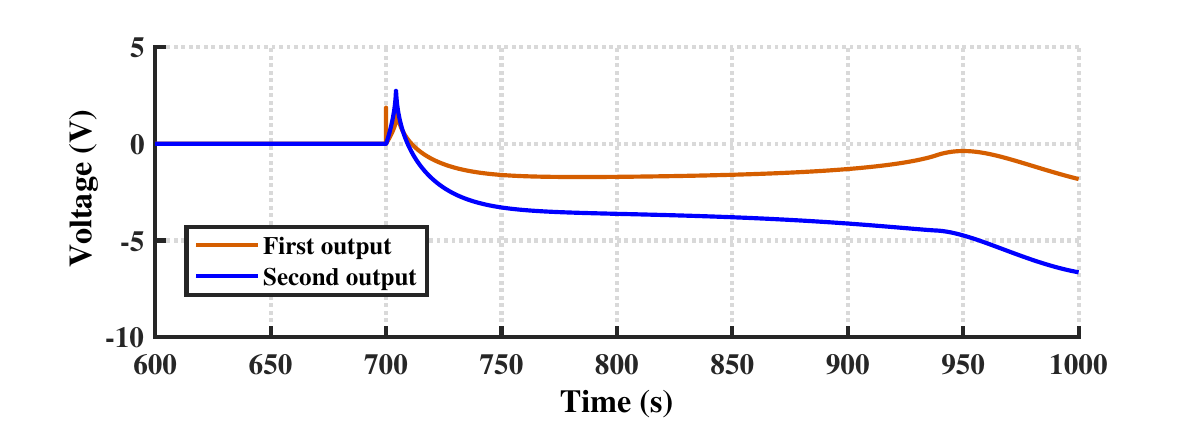}
	\caption{\textcolor{black}{The neural network output in scenario A: Recovery signals are used to restore the closed-loop system to its normal operating point.}}
	\label{fig::neural network}	
\end{figure}

\indent  The residual signals are shown in Fig.~(\ref{fig::attacknon}). When the attack occurs, the residual signals become nonzero. In contrast, residual signals converge to zero during steady states and in the absence of the attack signals.
 As expected, at $t = 700$~s, the residual signal becomes nonzero after the occurrence of the attack. Considering that the attack happens at $t = 700$~s, it can be concluded that the attack detection time is zero.
\\\indent The difference between the residual and attack signals is shown in Fig.~(\ref{fig::resnon}). As we can see, after the attack happens at $t = 700$~s, the difference between the residual and attack signals converges to zero.
\\\indent Zero dynamics of the system after applying the recovery signal and its absence are shown in Fig. (\ref{fig::z1}) and Fig. (\ref{fig::z2}). The attack does not change the zero dynamics of the system in the presence of the recovery algorithm. Without the recovery algorithm, the system's zero dynamics converge to zero, which means the tanks become empty. \textcolor{black}{The attack detection time and success rate for this scenario are zero, validating the recovery and detection algorithm's acceptable performance.}
\indent \textcolor{black}{The estimated recovery signals, which are outputs of the neural network are shown in Fig. \ref{fig::neural network}. Before the attack is detected, the neural network outputs are zero, and after attack detection, the neural network outputs converge to the attack signal.}
\subsection{\textcolor{black}{Performance Evaluation with Noisy Measurements} \label{noise}}
\begin{figure}[H]
	\centering
	\includegraphics[width=1\linewidth]{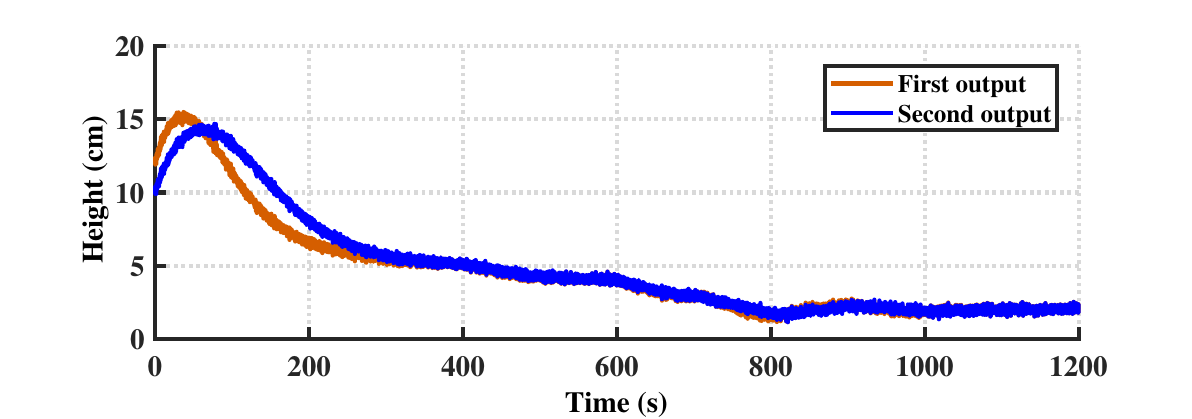}
	\caption{\textcolor{black}{System output in scenario B: The output of the closed-loop system under noisy measurements.}}
	\label{fig::outnoise}	
\end{figure}

\begin{figure}[H]
	\centering
	\includegraphics[width=1\linewidth]{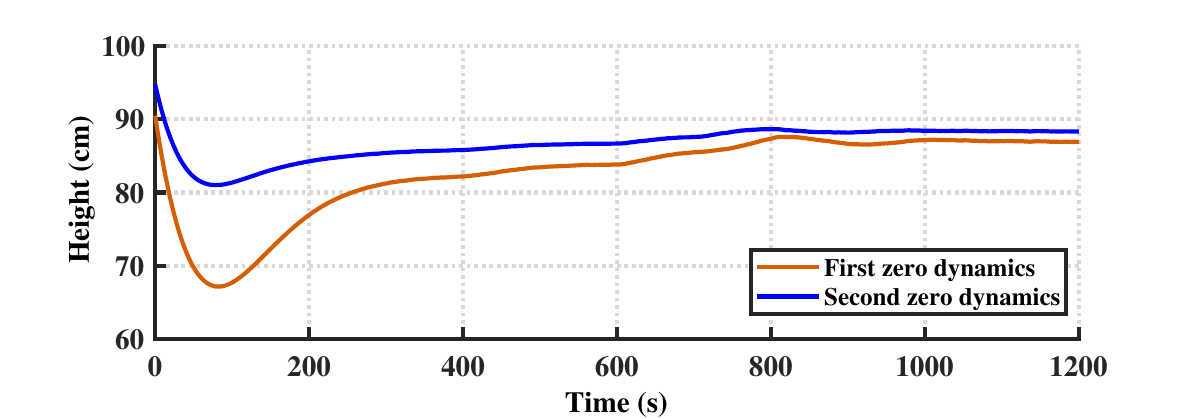}
	\caption{\textcolor{black}{Zero dynamics the system in scenario B: Zero dynamics of the system stay stable under the proposed recovery algorithm with noisy measurements.}}
	\label{fig::zrnoise}	
\end{figure}

\begin{figure}[H]
	\centering
	\includegraphics[width=1\linewidth]{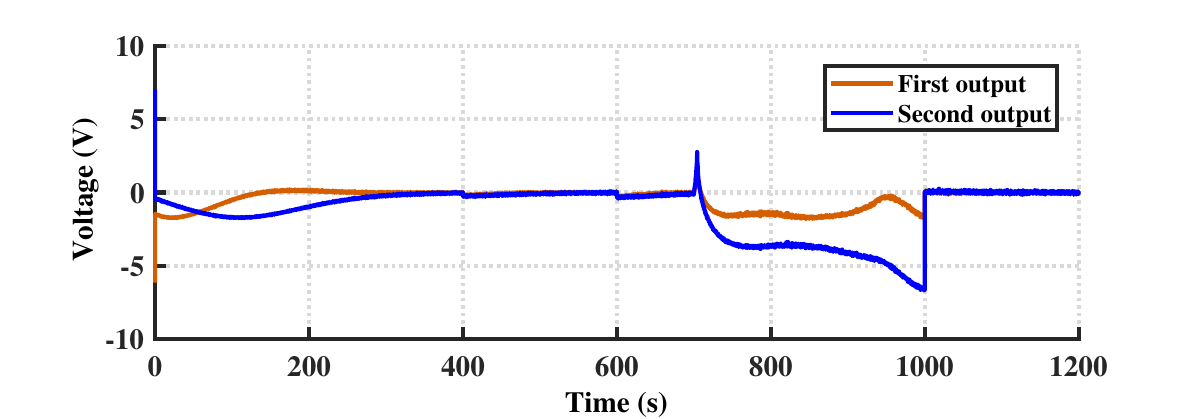}
	\caption{\textcolor{black}{Residual signals in scenario B: Residual signals used to detect the occurrence of the attack and update the neural network weights.}}
	\label{fig::rnoise}	
\end{figure}

\begin{figure}[H]
	\centering
	\includegraphics[width=1\linewidth]{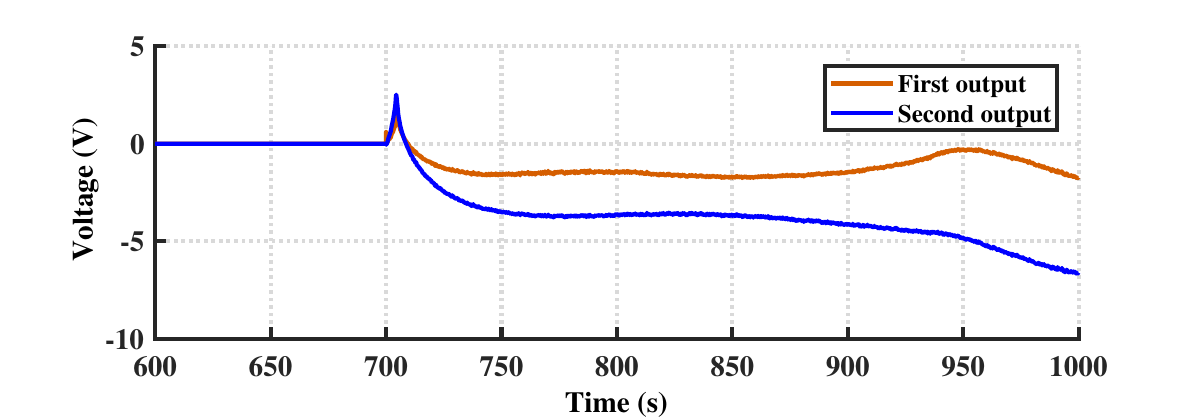}
	\caption{\textcolor{black}{The neural network output in scenario B: Recovery signals are used to restore the closed-loop system to its normal operating point irrespective of environmental disturbances.}}
	\label{fig::nnnoise}	
\end{figure}

 \textcolor{black}{In this scenario, measurements are corrupted by additive Gaussian noise with a noise power of $0.01$. Moreover, the reference signal is considered to change step wise from $5$ to $2$ to evaluate the performance of the algorithm with different operating points. More specifically, the reference signal is considered to be: 	
 	\begin{align}
y_{\text{ref}} = \begin{cases}
5 &0 \leq t \leq 300\\
3.7 & 300 < t \leq 500\\
2.7 &  500 < t \leq 705 \\
2  &\quad\text{otherwise.} \\ 
\end{cases}
 	\end{align}
 \indent The attack scenario is the same as the previous subsection. The numerical results  illustrated in Figs.~\ref{fig::outnoise} and \ref{fig::zrnoise} show the outputs and zero dynamics of the closed-loop system in the presence of the stochastic measurement noise. Moreover, the residuals signals are presented in Fig.~\ref{fig::rnoise}. In addition, the output of the neural network model, i.e., the recovery signal, is depicted in Fig.~\ref{fig::nnnoise}. We can see that irrespective of the stochastic disturbances, the presented strategy can restore the system under attack and guarantee the stability of the closed-loop system. It is worth mentioning that in this scenario the input of the neural network model is noisy as well, meaning that the proposed method is robust against imperfections and environmental disturbances. The attack detection time for this scenario is zero and the second index is $\gamma = 0.0041$. These indices are near zero and verify the proposed algorithms' efficiency with noisy measurements.}

\subsection{\textcolor{black}{Time-Varying Reference Signal} \label{timevar}}
\begin{figure}[H]
	\centering
	\includegraphics[width=1\linewidth]{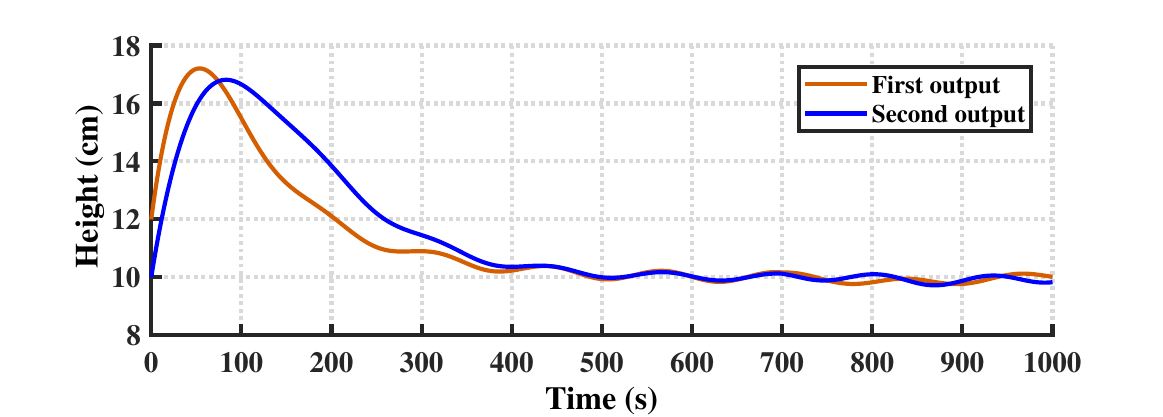}
	\caption{\textcolor{black}{System output in scenario C: The output of the closed-loop system with a time-varying reference signal.}}
	\label{fig::outvar}	
\end{figure}

\begin{figure}[h!]
	\centering
	\includegraphics[width=1\linewidth]{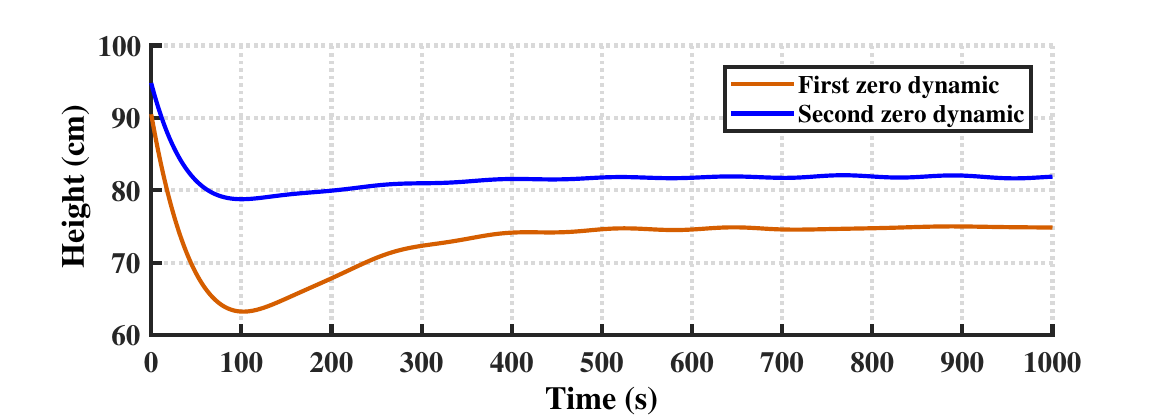}
	\caption{\textcolor{black}{Zero dynamics the system in scenario C: Zero dynamics of the system stay stable under the proposed recovery algorithm with a time-varying reference signal.}}
	\label{fig::zvar}	
\end{figure}

\begin{figure}[h!]
	\centering
	\includegraphics[width=1\linewidth]{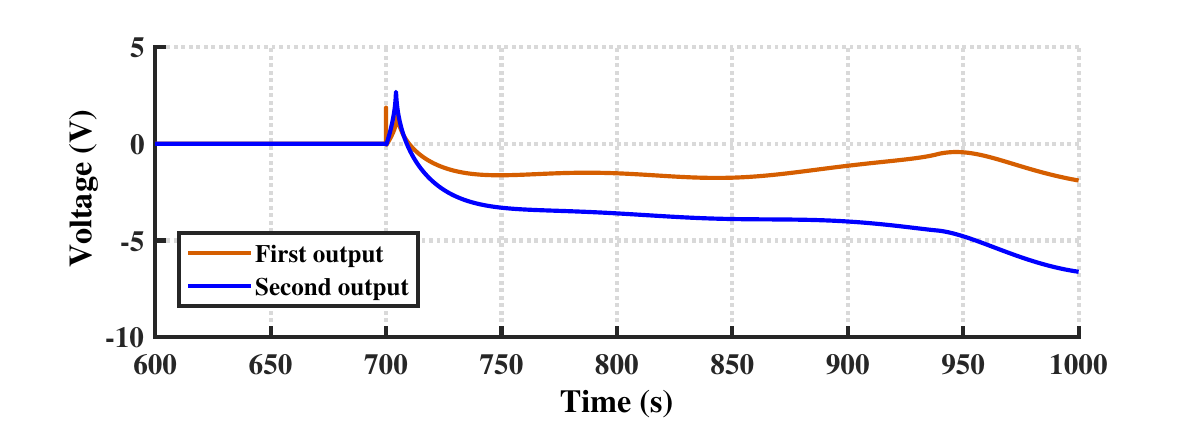}
	\caption{\textcolor{black}{The neural network output in scenario C: Recovery signals are used to restore the closed-loop system to its normal operating point irrespective of the properties of the reference signal.}}
	\label{fig::nnnvar}	
\end{figure}

\textcolor{black}{ In this subsection, the performance of the proposed strategy with a time-varying reference signal is evaluated. The reference signal is considered to be $10 + sin(0.05t)$. Zero dynamics and system outputs in the presence of the proposed recovery algorithm are shown in Figs.~\ref{fig::zvar} and ~\ref{fig::outvar}, respectively. The recovery signal in this scenario are presented in Fig.~\ref{fig::nnnvar}. Unlike the study done in \cite{Uncertain_nonlinear}, the proposed strategy does not impose any limitations on the properties of the reference signal. As a result, it can be seen that the neural network output converges to the attack signal with a bounded error. The recovery algorithm keeps the zero dynamics of the system stable, while the system output follows the reference signal flawlessly. Computing the performance indexes confirms the recovery and detection strategy's performance. The attack detection time and success rate are zero and $0.0019$, respectively. }
\subsection{Comparison Study\label{compare}}
\begin{figure}[H]
	\centering
	\includegraphics[width=1\linewidth]{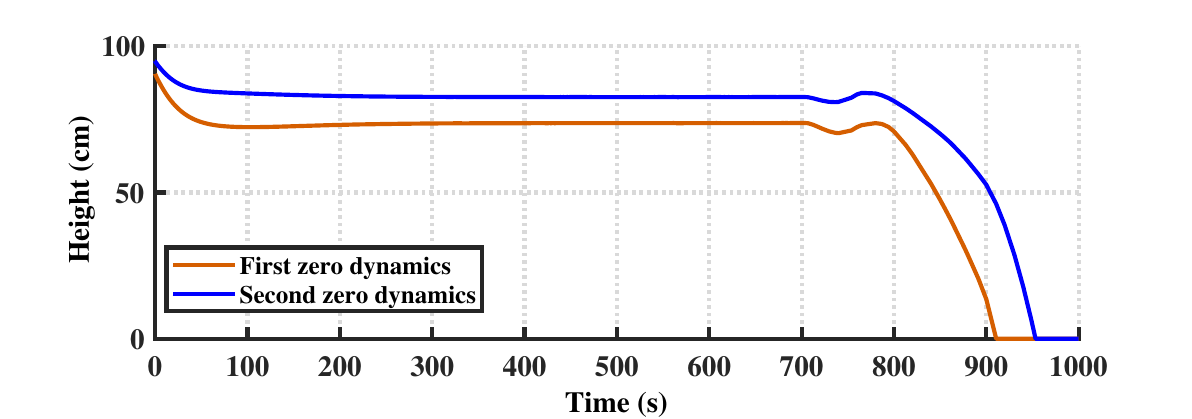}
	\caption{\textcolor{black}{Zero dynamics the system in scenario D: Zero dynamics of the system cannot show a desirable behavior under the proposed recovery algorithm in \cite{simulation-compare}.}}\label{fig::comparemethod}
\end{figure}

 In \cite{simulation-compare}, a resilient control strategy is presented to recover nonlinear affine systems under actuator attack. Fig.~ (\ref{fig::comparemethod}) shows the zero dynamics of the system in the presence of the recovery algorithm presented in \cite{simulation-compare}. The simulation scenario is the same as the scenario in subsection \ref{const}. As can be seen, this method cannot stabilize the zero dynamics of the system in the presence of a robust zero dynamics attack. More specifically, the attack success rate for the proposed method in \cite{simulation-compare} is almost $100 \%$, while in the proposed method, the attack success rate is zero.
\section{Conclusion \label{conclusion}}
\textcolor{black}{In this work, a new detection and stabilization strategy for general MIMO nonlinear cyber-physical systems subjected to robust zero dynamics attacks is presented. The detection stage uses a residual signal, and the recovery process is based on a neural network to estimate the attack signal. The proposed method ensures the stability of the closed-loop system and enables the system to return to its nominal operating point after being subjected to the attack. The strategy does not depend on knowing the exact model of the system, does not interfere with the system's controller, does not impact the system's performance, and is not computationally complicated, making it suitable to be implemented on typical microcontrollers in practical applications. The simulation results are used to highlight the performance of the strategy and verify the theoretical results. Future works include extending the obtained results to the systems subjected to packet loss and communication delays.}

\vspace{11pt}


\vspace{11pt}


\bibliographystyle{ieeetran}
\bibliography{IEEEabrv,root}

\vfill

\end{document}